\documentstyle[aps,epsf,prl,epsfig,twocolumn]{revtex}
\def\be{\begin{equation}}
\def\ee{\end{equation}}
\def\bea{\begin{eqnarray}}          
\def\eea{\end{eqnarray}}
\def\bi{\begin{itemize}}
\def\ei{\end{itemize}}

\begin{document}

\title{ Images of a Bose-Einstein condensate:            
        diagonal dynamical Bogoliubov vacuum  }

\author{ Jacek Dziarmaga and Krzysztof Sacha }

\address{ Institute of Physics and Centre for Complex Systems,
          Jagiellonian University,
          Reymonta 4, 30-059 Krak\'ow, Poland  }

\date{ March 4, 2005 }

\maketitle

\begin{abstract}
Evolution of a Bose-Einstein condensate subject to a time-dependent 
external perturbation can 
be described by a time-dependent Bogoliubov theory: a condensate 
initially in its ground
state Bogoliubov vacuum evolves into a time-dependent excited state 
which can be formally 
written as a time-dependent Bogoliubov vacuum annihilated by 
time-dependent quasiparticle 
annihilation operators. We prove that any Bogoliubov vacuum can be 
brought to a diagonal 
form in a time-dependent orthonormal basis. This diagonal form is 
tailored for simulations 
of quantum measurements on an excited condensate. As an example we work 
out phase imprinting
of a dark soliton followed by a density measurement. 
\end{abstract}

Quantum measurements on Bose-condensed systems can give quite 
unexpected results.
For example, in the classic paper by Javanainen and Yoo \cite{Fock} a 
density measurement
on a Fock state $|N/2,N/2\rangle$ with $N$ particles equally divided 
between two 
counter-propagating plane waves $e^{\pm ix}$ reveals an interference 
pattern 
$\rho(x|\varphi)\sim \cos^2(x-\varphi)$ with a phase $\varphi$ chosen 
randomly in every 
realization of the experiment. The Fock state has a uniform single 
particle density 
distribution, but its measurement unexpectedly reveals interference 
between the two 
counter-propagating condensates. The Fock state is a quantum 
superposition over 
$N$-particle condensates with different relative phases $\varphi$ in 
their wave functions 
\cite{phase}, 
$|N/2,N/2\rangle\sim\int 
d\varphi~|N:e^{+i(x-\varphi)}+e^{-i(x-\varphi)}\rangle$,
but every single realization of the experiment reveals such a density 
distribution as
if the state before the density measurement were one of the condensates 
$|N:e^{+i(x-\varphi)}+e^{-i(x-\varphi)}\rangle$ with a randomly chosen 
phase $\varphi$.
This effect is best explained \cite{phase} when the density 
measurement, which is a 
destructive measurement of all particle positions at the same time, is 
replaced by 
an equivalent sequential measurement of one position after another. 
With an increasing
number $n$ of measured positions a quantum state of the remaining $N-n$ 
particles
gradually ``collapses'' from the initial uniform superposition over all 
phases to a state
with a more and more localized phase $\varphi$. 
For a large $N$ a 
measurement of only a small fraction $\frac{n}{N}\ll 1$ of all 
particles practically 
collapses the state of remaining $N-n$ particles to 
a condensate with phase $\varphi$.
The lesson from this instructive example \cite{Fock,phase} is that when 
we want to predict 
possible outcomes of measurements on a Bose-condensed state, then it is 
helpful to rewrite 
the state as a superposition over condensates
$\int {\cal D}\phi~\psi(\phi)~|N:\phi\rangle$ with different condensate 
wave functions 
$\phi(\vec r)$ in a hope that $P(\phi)\sim|\psi(\phi)|^2$ will give an 
approximate probability 
for different density measurement outcomes $\rho(\vec 
r|\phi)=N|\phi(\vec r)|^2$. 
However, as the set 
of all condensates $|N:\phi\rangle$ is non-orthogonal and overcomplete, 
the ``amplitude''
$\psi(\phi)$ is not unique but depends on coordinates used to 
parameterize the space of 
condensate wave functions $\phi(\vec r)$. In general a special care 
should 
be taken to choose 
the right parameterization where the simple ``Born rule'' 
$P(\phi)\sim|\psi(\phi)|^2$ gives the 
correct probability.  

The existing literature about measurements on Bose-condensed systems 
\cite{Fock,phase} concentrates on simple most beautiful examples of 
Bose-condensed states. In this letter we construct for the first time 
a measurement theory able to predict probability of different density 
measurement outcomes on a realistic Bose-Einstein condensate of interacting 
particles evolving under influence of external time-dependent potentials. 
This problem is important because in many experiments, like phase imprinting
of dark solitons \cite{Hannover} or condensate splitting in atom interferometers 
\cite{interferometer}, manipulation of the condensate generates substantial 
dynamical depletion which can qualitatively affect measured density patterns.
In addition to being realistic our theory also reveals a beautiful diagonal
structure hidden in the quantum state of a condensate excited from its ground 
state by a time-dependent potential.

A generic Bose-Einstein condensate consisting of $N$ weakly interacting atoms 
can be described by a number-conserving version of the Bogoliubov theory 
\cite{CastinDum} which assumes that most particles occupy the condensate wave 
function. Time-dependent problems can be treated with the time-dependent version 
of the theory where the condensate wave function $\phi_0(t)$ evolves according 
to the time-dependent Gross-Pitaevskii equation (GPE) and the Bogoliubov modes 
$u_m(t)$ and $v_m(t)$ solve the time-dependent Bogoliubov-de Gennes equations 
(BdGE) \cite{CastinDum}. In this framework, starting with a condensate in the 
ground state, time-dependent perturbation excites the condensate to a state which 
is formally a time-dependent Bogoliubov vacuum $|0_b\rangle$ annihilated by the 
time-dependent quasiparticle annihilation operators
\be
\hat b_m(t)~=~
\frac{ \langle u_m(t)   | \hat\psi \rangle~         \hat a_0^\dagger -
       \langle v_m^*(t) | \hat\psi \rangle^\dagger~ \hat a_0           
}{\sqrt{N}}.
\ee
Here the operator $\hat a_0=\langle\phi_0(t)|\hat\psi\rangle$ annihilates 
an atom in a condensate wave function $\phi_0(t)$. The modes $u_m(t)$ and 
$v_m^*(t)$ are solutions of the time-dependent BdGE projected on the subspace 
orthogonal to $\phi_0$, i.e. 
$\langle u_m|\phi_0\rangle=\langle v_m^*|\phi_0\rangle=0$, and normalized so
that $\langle u_m|u_n\rangle-\langle v_m|v_n\rangle=\delta_{mn}$. In the 
time-dependent vacuum $|0_b\rangle$ the reduced single particle 
density matrix 
\be
\rho(\vec r,\vec r\;')=
N\phi_0^*(\vec r)\phi_0(\vec r~')+
\sum_{m=1}^\infty v_m(\vec r) v_m^*(\vec r~'),
\label{sing}
\ee
has common eigenstates with the operator
\be
{\rm d}\hat\rho=\sum_{m=1}^\infty|v_m\rangle\langle v_m |=
\sum_\alpha {\rm d}N_\alpha
|\phi_\alpha^*\rangle\langle\phi_\alpha^*|~.
\label{drho}
\ee
A mode $|\phi_\alpha^*\rangle$ is a non-condensate eigenmode of the density 
matrix (\ref{sing}) occupied on average by ${\rm d}N_\alpha$ particles. Thanks 
to the completeness relation
\be
\sum_m |u_m\rangle\langle u_m|-\sum_m |v_m^*\rangle\langle v_m^*|=1,
\label{complet}
\ee
valid in the subspace orthogonal to $|\phi_0\rangle$, the operator 
${\rm d}\hat\rho^*=\sum_m|v_m^*\rangle\langle v_m^*|$ has common 
eigenstates with the 
\be
\sum_{m=1}^\infty|u_m\rangle\langle u_m|=
1+{\rm d}\hat\rho^*=
\sum_\alpha
\left(1+{\rm d}N_\alpha\right) |\phi_\alpha\rangle\langle\phi_\alpha|.
\ee
Furthermore $|\phi_\alpha\rangle$'s allow us to get a diagonal form of 
\be
\hat\Delta=\sum_m|v_m\rangle\langle u_m|=
\sum_\alpha
\Delta_\alpha
|\phi_\alpha^*\rangle\langle\phi_\alpha|,
\label{hatDelta}
\ee
which is a map between the subspace orthogonal to $|\phi_0\rangle$ and
the subspace orthogonal to $|\phi_0^*\rangle$. Indeed, the property
$\langle u_m|u_n\rangle-\langle v_m |v_n\rangle=\delta_{nm}$ of the 
Bogoliubov modes and Eq.(\ref{complet}) imply a quasi-commutator
${\rm d}\hat\rho\hat\Delta=\hat\Delta{\rm d}\hat\rho^*$
from which follows the diagonal form in Eq.(\ref{hatDelta}).
What is more, owing to
${\rm d}\hat\rho^*(1+{\rm d}\hat\rho^*)=\hat\Delta^*\hat\Delta$,
one gets 
$
|\Delta_\alpha|^2={\rm d}N_\alpha(1+{\rm d}N_\alpha).
$
In Ref.~\cite{stanz} we have shown that in the stationary case the
Bogoliubov vacuum has the following form 
\be
|0_b\rangle~\sim~
\left(
\hat a_0^\dagger\hat a_0^\dagger +
\sum_{\alpha,\beta=1}^\infty
Z_{\alpha,\beta}
\hat{\tilde a}_\alpha^\dagger \hat{\tilde a}_{\beta}^\dagger 
\right)^{N/2}~
|0\rangle~,
\label{nz}
\ee
where $\hat{\tilde a}_\alpha$'s annihilate atoms in single particle 
states $\tilde\phi_\alpha$ orthogonal to the condensate wave function 
$\phi_0$. Matrix $Z_{\alpha,\beta}$ fulfills the equation
\be
\langle v_m|\tilde\phi_\alpha^*\rangle=\langle u_m|\tilde 
\phi_{\beta}\rangle 
Z_{\beta,\alpha}, 
\label{Z}
\ee
which is equivalent to the set of the conditions $\hat b_m|0_b\rangle=0$. In 
the stationary and time reversal invariant case all modes can be chosen as real 
functions so that the matrix $Z_{\beta,\alpha}$ is real symmetric and can be 
diagonalized by an orthogonal transformation \cite{stanz,leggett}. In a generic 
time-dependent case, or when a stationary $\phi_0$ breaks the $T$-invariance 
like in e.g. vortex state, the matrix $Z$ is complex and symmetric. However, 
using only the general properties of ${\rm d}\hat\rho$ and $\hat\Delta$ it is 
easy to show that the choice of $|\phi_\alpha\rangle$'s for the modes 
$|\tilde\phi_\alpha\rangle$ in Eq.(\ref{nz}) makes the time-dependent Bogoliubov 
vacuum diagonal 
\be
|0_b\rangle~\sim~
\left(
\hat a_0^\dagger \hat a_0^\dagger+
\sum_{\alpha=1}^\infty
\lambda_\alpha
\hat a_\alpha^\dagger \hat a_\alpha^\dagger
\right)^{N/2}~
|0\rangle~,
\label{0lambda}
\ee
where the eigenvalues of the matrix $Z$ are
\be
\lambda_\alpha=\frac{{\rm d}N_\alpha}{\Delta_\alpha}.
\label{lam}
\ee
Indeed, an action of $\sum_m\langle \phi_\alpha^*|v_m\rangle$ on both 
sides of Eq.~(\ref{Z}) yields
$
\langle \phi_\alpha^*|{\rm d}\hat\rho|\phi^*_{\beta}\rangle=
\langle \phi_\alpha^*|\hat\Delta|\phi_{\alpha'}\rangle 
Z_{\alpha',\beta}~,
$
equivalent to Eq.(\ref{lam}).

Hence the procedure of time evolution of the Bogoliubov vacuum state in 
the particle 
representation becomes extremely simple: i) First one has to evolve 
Bogoliubov 
modes $(u_m,v_m)$ together with the condensate wave function $\phi_0$ 
which 
is easy because different modes evolve independently from each other. 
ii) Next one has to diagonalize the operator 
${\rm d}\hat\rho=\sum_m|v_m\rangle\langle v_m|$ (diagonalization of 
$1+{\rm d}\hat\rho^*=\sum_m|u_m\rangle\langle u_m|$ turns out to be 
better convergent numerically) in order to get eigenvalues ${\rm d}N_\alpha$ 
and eigenvectors $|\phi_\alpha\rangle$. iii) Finally one calculates
$\Delta_\alpha=\langle\phi_\alpha^*|\hat\Delta|\phi_{\alpha}\rangle$ 
and from Eq.~(\ref{lam}) obtains values of $\lambda_\alpha$. Note that 
there is a one to one correspondence between moduli of $\lambda_\alpha$ 
and the eigenvalues of the single particle density matrix ${\rm d}N_\alpha$:
\be
|\lambda_\alpha|=\frac{{\rm 
d}N_\alpha}{|\Delta_\alpha|}=\sqrt{\frac{{\rm
d}N_\alpha}{{\rm d}N_\alpha+1}}.
\ee
However, the phases of $\lambda$'s cannot be determined from the single 
particle matrix 
only, but they also depend on the eigenvalues
$\Delta_\alpha=\langle\phi_\alpha^*|\hat\Delta|\phi_\alpha\rangle$, see 
Eq.(\ref{lam}). 
Finally, once the phases are known it is convenient to make the 
transformation
\be
\lambda'_\alpha = |\lambda_\alpha|~,~~  
\phi'_\alpha    = \phi_\alpha e^{\frac12i{\rm Arg}(\lambda_\alpha)}~.
\ee
In the following we skip the primes.

For large $N$ the diagonal state (\ref{0lambda}) can be rewritten 
as a superposition over $N$-particle condensates
\begin{eqnarray}
|0_b\rangle~\sim~
\int dq~
e^{-\sum_{\alpha}
    \frac{1-\lambda_\alpha}{2\lambda_\alpha}
    q_\alpha^2 }~
|N:\phi(\vec r|q)\rangle~.
\label{0q}
\end{eqnarray}
with the normalized condensate wave functions
\begin{equation}
\phi(\vec r|q)~=~
\frac{ \phi_0(\vec r)+\frac{1}{\sqrt{N}}\sum_\alpha 
q_\alpha\phi_\alpha(\vec r) }
     { \sqrt{1+\frac{1}{N}\sum_\alpha q_\alpha^2} }~.
\label{phiq}
\end{equation}
The gaussian amplitude gives an accurate vacuum state when 
$\sum_\alpha {\rm d}N_\alpha \ll \sqrt{N}$.
The gaussian square of the amplitude 
\be
P(q)\sim 
\prod_\alpha e^{-\frac{1-\lambda_\alpha}{\lambda_\alpha}q_\alpha^2}
\label{Pq}
\ee 
is a candidate for the probability distribution for different $q_\alpha$. 
If this choice is correct, then simulation of the density measurement amounts 
to choosing a random set of $q_\alpha$'s which determines the measured density 
\be
\rho(\vec r|q)=
N|\phi(\vec r|q)|^2~.
\label{rhoq}
\ee
To show the correctness of Eq.(\ref{Pq}), we average $\rho(\vec r|q)$ 
over $P(q)$ and for $\sum_\alpha {\rm d}N_\alpha\ll N$ we get the density
\be
\rho(\vec r)=
N|\phi_0(\vec r)|^2+
\sum_\alpha \frac{\lambda_\alpha}{2(1-\lambda_\alpha)} 
|\phi_\alpha(\vec r)|^2
\label{rhoav}
\ee
averaged over different realizations of $q$. This average density should be equal 
to the expectation value of the density operator in the state (\ref{0q})
\be
\langle 0_b|\psi^\dagger(\vec r)\psi(\vec r)|0_b\rangle=
N|\phi_0(\vec r)|^2+
\sum_\alpha {\rm d}N_\alpha |\phi_\alpha(\vec r)|^2
\label{rhoexp}
\ee
with ${\rm d}N_\alpha=\lambda_\alpha^2/(1-\lambda_\alpha^2)$. These two densities 
are the same for the highly occupied modes with ${\rm d}N_\alpha\gg 1$ or 
$\lambda_\alpha\approx 1^-$. There are discrepancies for poorly occupied modes 
but those modes give negligible contribution to the total density. The probability
(\ref{Pq}) accurately reproduces the single particle density matrix. This accuracy
can be best explained when we look at the overlap between condensates
\be
\langle N:\phi(\vec r|q)|N:\phi(\vec r|q')\rangle=
e^{-\sum_\alpha(q_\alpha-q'_\alpha)^2}~.
\ee  
Condensates become orthogonal on the length scale of $1$. For the highly occupied
modes the fluctuations of $q_\alpha\sim\sqrt{{\rm d}N_\alpha}$ are much greater 
than the width of the overlap and for these modes the overlap can be replaced by 
a delta function $\delta(q_\alpha-q'_\alpha)$. With the delta overlap the 
condensates are orthogonal and the probability $P(q)$ is simply given by the
Born rule in Eq.(\ref{Pq}).

To summarize, a density measurement, which is a measurement of all 
atomic positions, 
can be approximately simulated in two steps. In the first
step a condensate wave function (\ref{phiq}) is chosen randomly from 
the gaussian distribution 
$P(q)\sim\prod_\alpha \exp(-q_\alpha^2/2{\rm d}N_\alpha)$. In the 
second step atomic 
positions are measured in the chosen condensate. The first step already 
gives a smoothed 
density distribution $N|\phi(\vec r|q)|^2$ on top of which the second 
step 
only 
adds statistical fluctuations. In most applications one is not 
interested in the 
statistical fluctuations but in the smoothed density obtained after 
filtering out the 
statistical noise, compare the histogram and the solid line in Fig.2. 
Thus for most 
applications the density measurement can be very efficiently and 
accurately simulated with 
only the first step of the procedure which immediately gives a smoothed 
density 
$\rho(\vec r|q)=N|\phi(\vec r|q)|^2$ with randomly chosen $q$'s. In the 
following 
we give an example 
of this procedure: density measurement after phase imprinting of a dark 
soliton.

\begin{figure}
\centering
\includegraphics*[width=8.6cm]{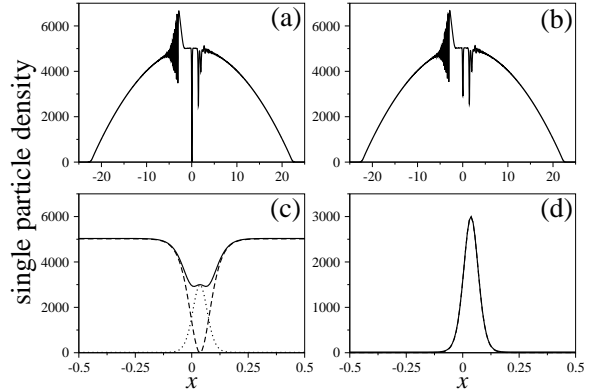}
\caption{  
Panel (a) shows condensate density, i.e. $N|\phi_0(x,t)|^2$, at the time 
$t=0.15$. In panel (b) single particle density 
$\rho(x,t)=N|\phi_0(x,t)|^2+\sum_m |v_m(x,t)|^2$ is plotted. Panel (c) shows 
single particle density 
$\rho(x,t)$ (solid line), condensate density $N|\phi_0(x,t)|^2$ (dashed 
line) and density of 
noncondensate atoms $\sum_m |v_m(x,t)|^2$ (dotted line) in the 
vicinity of the trap center.
In panel (d) $\sum_m |v_m(x,t)|^2$ (solid line) is plotted together 
with 
${\rm d}N_1|\phi_1(x,t)|^2$ (dashed line barely visible behind the 
solid line) where 
$\phi_1(x,t)$ is the most occupied mode (${\rm d}N_1\approx273$) 
orthogonal to 
$\phi_0(x,t)$. The parameters correspond to the Hannover experiment
where $N=1.5\cdot 10^5$ (time $t=0.15$ corresponds to 1.7~ms).
}
\label{one}
\end{figure}

To describe evolution of the condensate wave function we solve the 
time-dependent GPE that in harmonic oscillator trap units 
reads
\be
i\partial_t\phi_0=
-\frac12\partial_x^2\phi_0+
\frac{x^2}{2}\phi_0+
g|\phi_0|^2\phi_0+
V(t,x)\phi_0~.
\label{GPE}
\ee
Here $V(t,x)$ is an external potential created by the laser beam. As 
the effective 1D interaction strength we choose $g=7500$ corresponding to 
parameters of the Hannover experiment \cite{Hannover}. The dark soliton is 
a kink $\phi_0\sim\tanh(x/\xi)$ with the wave function winding a phase of 
$\pi$ as $x$ goes from negative to positive. At the same time the density 
drops to zero at $x=0$ explaining the name dark soliton. Experimentally the 
soliton is excited by a short laser pulse which imprints on the wave function 
a phase that changes by $\pi$ when one goes from negative to positive $x$. 
To simulate the phase imprinting we started with the ground state of the 
stationary GPE for a condensate in the harmonic trap, $\phi_0(x,0)$, and applied 
(similarly as in Ref.\cite{Law} where they use rather small $g=200$) the 
perturbation $V(t,x)=V_0[1+\tanh(x/l_0)]$ (where $V_0=140$, $l_0=0.05$) that 
lasted for $0.034$. Then $\phi_0(x,t)$ was evolved up to $t=0.15$ (or $1.7$ms). 
The density of the resulting wave function is plotted in Fig~\ref{one}(a) where 
the dark soliton is clearly visible in the center of the trap.

The Bogoliubov modes solve the time-dependent BdGE
\bea
i\partial_t u_m&=&\left(H_{\rm GP}+g|\phi_0(x,t)|^2\right) u_m +
g\phi_0^2 v_m,   \cr
i\partial_t v_m&=&-\left(H_{\rm GP}+g|\phi_0(x,t)|^2\right)v_m -
g{\phi_0^*}^2 u_m,
\eea
where $H_{\rm GP}=-\partial_x^2/2 +x^2/2+ V(x,t)+g|\phi_0(x,t)|^2$. We 
have evolved the Bogoliubov modes, starting with the eigenmodes of the 
stationary BdGE for a condensate in the ground state, up to $t=0.15$. 
The single particle density $\rho(x,t=0.15)$ is 
plotted in Fig.~\ref{one}(b). In Fig.~\ref{one}(c) the densities of condensate, 
noncondensate atoms and the total single particle density are plotted. 
At both values of $g$ (the realistic $g=7500$ here and the $g=200$ in 
Ref.\cite{Law}) the single particle density $\rho(x,t)$ shows that minimum at 
the soliton location is gradually filled with particles depleted from the 
condensate. These dynamical studies confirm earlier predictions that a 
static soliton is going to fill up after a few milliseconds \cite{greying}.   

\begin{figure}
\centering
\includegraphics*[width=8.6cm]{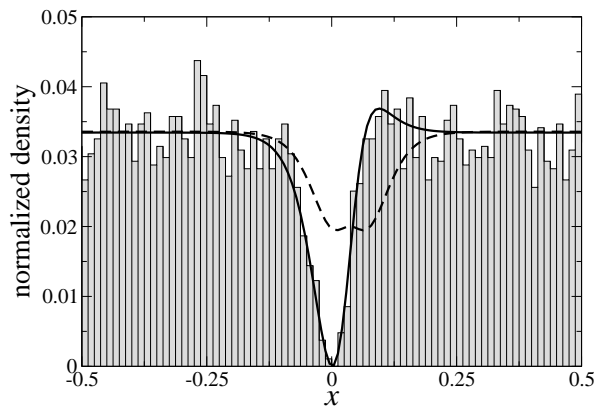}
\caption{ The histogram is the normalized
density distribution measured in a single exact 
simulation of the density measurement. The solid line is a fit to the histogram 
with one of the predicted $\rho(x|q_1)=N|\phi_0(x)+q_1\phi_1(x)/\sqrt{N}|^2$ with 
$q_1=14.5$. The dashed line is the total average density $\rho(x)$ which is equal 
to an average over densities measured in many experiments. }
\label{two}
\end{figure}

After the evolution of the Bogoliubov modes was completed we calculated 
the eigenvalues $\lambda_\alpha$ and the non-condensate eigenmodes 
$\phi_\alpha(x)$ at the final time. It turns out that the soliton notch is 
filled up with atoms depleted to only one mode: the $\phi_1(x)$ with the 
largest ${\rm d}N_1=273$.  In Fig.\ref{two}(d) the total depletion density 
$\sum_m|v_m(x)|^2$ (solid line) is plotted together with a barely visible 
dashed plot of the ${\rm d}N_1|\phi_1(x)|^2$. As the depletion in the soliton 
notch comes only from $\alpha=1$ and because we would like to know what will 
be seen in the soliton notch in a single experiment we truncate the final 
time-dependent Bogoliubov vacuum to 
$\left( \hat a_0^\dagger\hat a_0^\dagger +
        \lambda_1
        \hat a_1^\dagger\hat a_1^\dagger   \right)^{N/2}~|0\rangle$.
We made exact simulation of density measurement in the truncated vacuum 
using the algorithm of Ref.\cite{Fock} and Ref.\cite{greying}. A typical outcome 
is shown as the histogram in Fig.\ref{two}. It turns out that every observed 
histogram can be very well fitted with the density
$\rho(x|q_1)=N|\phi_0(x)+q_1\phi_1(x)/\sqrt{N}|^2$ where the only free 
parameter is $q_1$. The histogram in Fig.\ref{two} is fitted by the density 
$\rho(x|14.5)$ shown in the figure as the solid line. This is in agreement with 
our measurement theory predicting the smoothed histograms to be $\rho(x|q_1)$ 
with a random $q_1$. 


In conclusion, we derived a convenient diagonal form of the time-dependent
Bogoliubov vacuum which greatly facilitates simulations of quantum measurements
on Bose-condensed systems.

\section*{ Acknowledgements } 
We thank Zbyszek Karkuszewski for stimulating discussions. KS was 
supported by the KBN grant PBZ-MIN-008/P03/2003. The work of JD
was supported by Polish Goverment scientific funds (2005-2008) as a 
research project.


\end{document}